\documentclass[aps,prl,preprint,showkeys,showpacs]{revtex4}
\usepackage{graphicx}
\begin{document}

\title{First principles determination of the Peierls stress of the shuffle screw dislocation in silicon}

\author{L. PIZZAGALLI}
\email{Laurent.Pizzagalli@univ-poitiers.fr}
\author{P. BEAUCHAMP}
\affiliation{Laboratoire de M\'etallurgie Physique\\
CNRS UMR 6630, Universit\'e de Poitiers, \\
B.P. 30179 \\
86962 Futuroscope Chasseneuil Cedex, France}

\date{\today}

\begin{abstract}

The Peierls stress of the $a/2 \langle 110 \rangle$ screw dislocation belonging
to the shuffle set is calculated for silicon using density functional theory. 
We have checked the effect of boundary conditions by using  two models, the
supercell method where one considers a periodic array of dislocations, and the
cluster method where a single dislocation is embedded  in a small cluster. The
Peierls stress is underestimated with the supercell and overestimated with the
cluster. These contributions have been calculated and the Peierls stress is
determined  in the range between  $2.4\times10^{-2}$ and
$2.8\times10^{-2}$~eV~\AA$^{-3}$. When moving, the dislocation follows the
\{111\} plane going through a low  energy metastable configuration and never
follows the ${100}$ plane, which includes a higher energy metastable core
configuration.

\end{abstract}


\maketitle

\subsection{1. Introduction} \label{intro}

\noindent The Peierls stress is defined as the minimal stress needed to move a dislocation
from one lattice site to the next one at 0~K \cite{Hir82WIL}. It is a
fundamental   quantity in the study of plasticity in materials science.
Close-packed metals, ductile, have low Peierls stresses, typically 10$^{-4}\mu$
\cite{notemu}, whereas values 2 to 3 orders of magnitude higher are associated
with brittle covalent crystals. Experimentally, its determination is difficult
since  materials are brittle at low temperature. Then it is usually obtained by
extrapolating measured yield strengths to the absolute zero temperature.  In
addition, several Peierls stresses can be defined for a given material, one for
each kind of dislocation, and the separation of the different contributions  in
experiments is not an easy task.

The measurement is especially difficult in covalent crystals, since the brittle
to ductile transition happens at high temperature, leading to rough  estimation
of the Peierls stress. The case of silicon, which is often considered as a model
material for the study of cubic  diamond and zinc blende covalent crystals, and,
as a result, has been extensively studied, is a good illustration. Experiments
indicate a value ranging between  0.1$\mu$ and 0.5$\mu$ \cite{expeSP}, obtained
from yield strength measurements in the ductile regime. For these temperatures,
it is known that dislocations are  dissociated and glide along the narrowly
spaced ${111}$ planes \cite{Ale86NAB}, thus called ``glide planes''
\cite{Hir82WIL}. It has been suggested that  these partial dislocations move by
nucleation and propagation of kink pairs \cite{Due96PML,Bul01PMA}. The
experimental Peierls value is then associated with  the displacement of
partials.

However, new information on the plasticity of silicon and other diamond cubic
materials has recently emerged, concerning  deformation at 
unusually low temperatures. Experimentally, silicon samples submitted to large shear stress
components, have been deformed at low temperature without failure. This has been
made  possible by applying a very high confining pressure or in scratch tests
\cite{Rab00JPCM, Rab01MSE, Rab01SM}. The observed dislocation configuration
differs strongly from the usually reported dissociated dislocations seen after
deformation at higher temperature. The dislocations are undissociated,  and thus
most likely glide along the widely spaced ``shuffle planes'' \cite{Hir82WIL},
and are  aligned along several favoured orientations, $\langle110\rangle$/screw,
$\langle112\rangle/30^\circ$ and $\langle123\rangle/41^\circ$.  The authors
pointed out \cite{Rab00PSS} that the observation  of the non dissociated
dislocations fits into the analysis of Duesbery and Joos, which, if
extrapolated, predicts a transition from dissociated dislocations  in the glide
set in the high temperature/low stress domain, to undissociated dislocations in
the shuffle set in the low temperature/large stress domain.  Similarly,  in
III-V compounds with the zinc-blende structure, deformation  experiments at low
temperature under high confining pressure, indicate that the low temperature
plastic deformation is governed by  undissociated screw dislocations
\cite{Suz98PML}. It is then very important to characterize shuffle dislocations
to understand the  plasticity properties of these materials.

Despite the increasing interest, there are few quantitative informations on the
energetics and mobility properties of the dislocations of the shuffle set,
compared to those  already obtained for the partials in the glide set, which
have been the subject of most studies until recently \cite{Bul01PMA}.  For
example, as far as we know, no attempts have been made to determine 
experimentally the Peierls stress of perfect dislocations in cubic diamond or
zinc  blende materials. However, several calculations have been performed for
addressing these issues. Naturally, the screw orientation is selected   because,
among all characters, it always plays a special role. Indeed, it allows  for
cross-slip, and in the particular case of the diamond cubic structure, it is the
orientation where transition from glide set to shuffle set is possible  without
any diffusion. Moreover, in the observations after low temperature deformation,
it appears as one of the favoured orientations, probably indicative of
significant Peierls valleys \cite{Rab01MSE}.

The screw dislocation in the shuffle set has been the subject of recent investigations, 
such as stability calculations using empirical interatomic potentials \cite{Koi00PMA} and ab initio methods \cite{Ari94PRL,Piz03PMA}. 
Without going back to early elasticity based estimates of the energetics of the screw dislocation \cite{Cel61JPCS,Tei67PSS},
the Peierls stress on the screw dislocation in the shuffle set has been estimated to be 0.048~eV~\AA$^{-3}$ by Duesbery and al \cite{Joo94PRB}, 
with the Peierls-Nabarro model. More direct atomistic calculations with the Stillinger-Weber potential by two different groups led to 
0.036~eV~\AA$^{-3}$ \cite{Ren95PRB} and 0.013~eV~\AA$^{-3}$ \cite{Koi00PMA}. Miyata et al performed ab initio calculations and deduced 
somewhat indirectly a Peierls stress ranging from 0.14 to 0.19~eV~\AA$^{-3}$ \cite{Miy01PRB}, that is about 15 times the value proposed in \cite{Koi00PMA}. 
However, these approaches suffer from some insufficiencies. If the Peierls-Nabarro model is shown to represent well planar dislocation 
cores \cite{Ren95PRB}, the predicted values for the Peierls stress are more doubtful since this model considers the dislocation core 
as an undeformable object moving across the atomic rows. Concerning the use of the Stillinger-Weber potential, it has been 
shown \cite{Piz03PMA}, that it does not predict the stability of the various screw core configurations which is a necessary 
prerequisite to any reliable Peierls stress determination. It is also necessary to have a quantitative estimate of the effect 
of the boundary conditions, particularly in the relatively small cell sizes used in ab initio computations.

In this paper, we present the Peierls stress calculation of the shuffle screw dislocation using an ab initio method. In the 
same spirit of a previous report on the structure and energetic of this dislocation \cite{Piz03PMA}, the boundary effects 
which cannot be avoided in this kind of calculations were carefully determined. The Peierls stress has been calculated for 
both periodic and fixed conditions, and has been found in the range between 0.024 and 0.028~eV \AA$^{-3}$ when the dislocation 
moved in the $\langle112\rangle$ direction. This direction has been found to be the easiest, compared to the $\langle110\rangle$ direction.

\subsection{2. Computational methods}

\begin{figure}[ht]
\includegraphics[width=8.6cm]{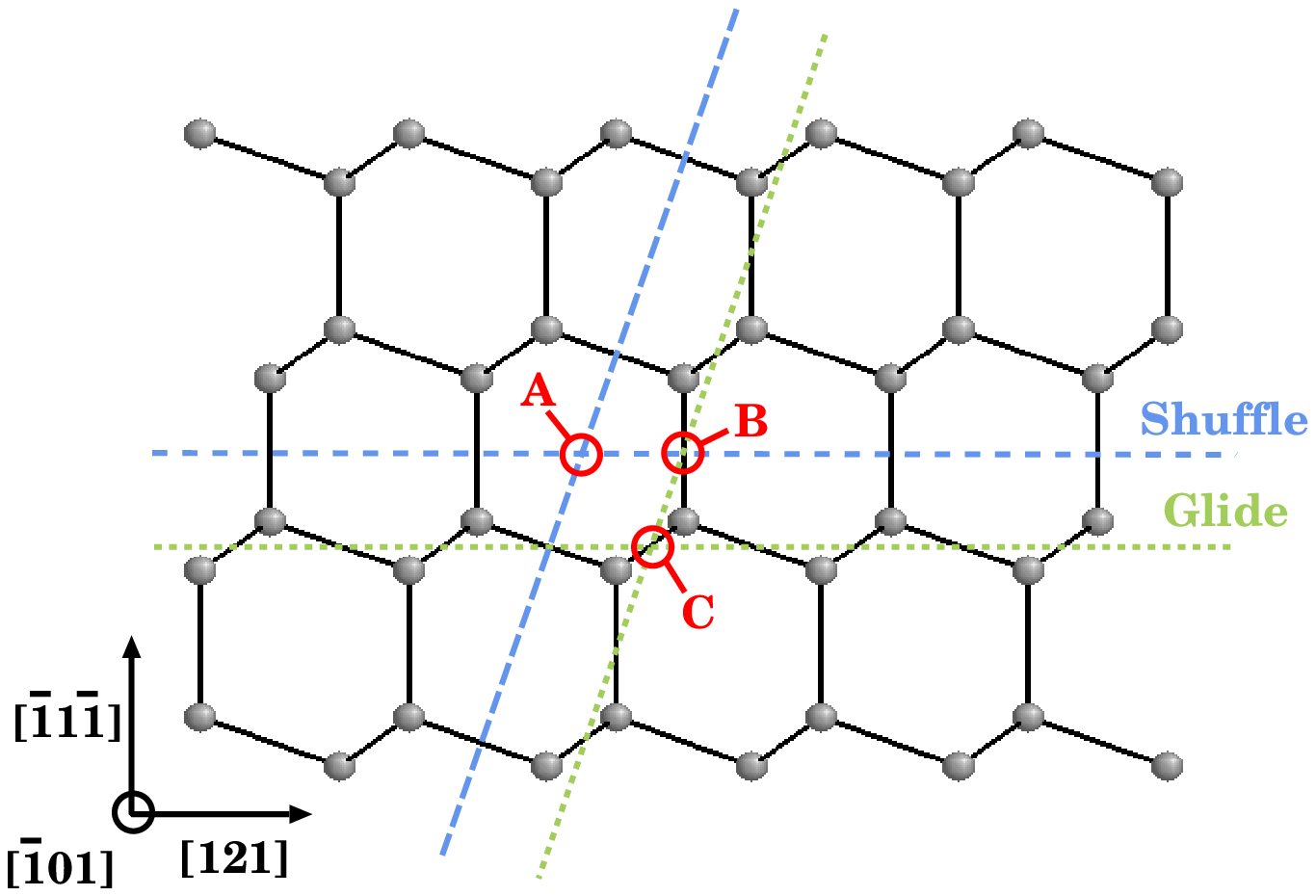}
\caption{Ball-and-stick representation of the cubic diamond structure, oriented
along axis $X=[121]$, $Y=[\bar{1}1\bar{1}]$ and $Z=[\bar{1}01]$.
Shuffle and glide sets of planes are shown, as well as the three possible stable
locations A, B, and C
for a screw dislocation in silicon \cite{Piz03PMA}. }\label{Shuffleglide}
\end{figure}

\noindent The determination of the Peierls stress is done by increasingly shearing the system, oriented as shown in the Figure~\ref{Shuffleglide}, 
so that the force on the dislocation is maximum along a chosen direction. Between each shear strain increment of 1\%, the atomic positions 
are relaxed. The corresponding Peierls stress is reached when the dislocation moves. Two selected directions, $\langle112\rangle$ 
and $\langle110\rangle$, have been investigated in this work. Starting from the most stable configuration A \cite{configscrew}, the 
first direction corresponds to a sequence ABA, whereas the second one should favour a sequence ACA (Fig.~\ref{Shuffleglide}).

The simulations have been performed in the framework of density functional
theory \cite{DFT} and local density approximation \cite{LDA}, by using
the code ABINIT \cite{ABINIT}. The ionic interactions were described by norm-conserving Troullier-Martins pseudopotentials \cite{Pseudo}. We used
a plane wave energy cutoff of 10~Ry and 2~special k-points along the dislocation
line. Previous calculations with similar parameters have allowed us to
model accurately screw dislocations in silicon \cite{Piz03PMA}.

\begin{figure}[ht]
\includegraphics[width=8.6cm]{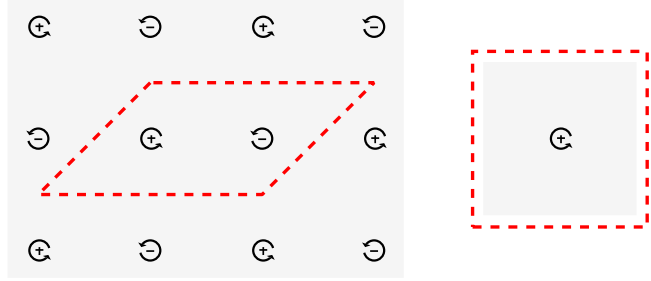}
\caption{Models used in the simulations: Oblique cell including a dislocations dipole
with periodic boundary conditions (left), and cell with a single dislocation and
fixed boundary conditions (right).}\label{CondLim}
\end{figure}

Two crystal models have been constructed \cite{Leh99EMIS}. In the first one, a dislocation dipole, cut from an infinite quadrupolar distribution, 
is placed in the crystal unit cell, and periodic boundary conditions are used in both directions normal to the dislocation line 
(Fig.~\ref{CondLim}). In the second model, a single dislocation is placed at the centre of a cluster whose lateral surfaces are kept fixed. 
In both cases, periodic boundary conditions are applied along the dislocation line. We have considered these two models because, as discussed in 
the following, they modify differently the determined Peierls stress. In the fully periodic case the interaction of the dislocation with the 
surrounding ones helps its move, so tending to underestimate the Peierls stress, whereas in the cluster model, the interaction with the 
fixed boundaries opposes the dislocation displacement and leads to an overestimation of the Peierls stress. Besides these purely mechanical effects, 
the two methods show also differences in the electronic structure whose effects become unimportant for cells or clusters large enough.

In case of periodic boundary conditions, we used a $12\times12\times1$ cell, i.e. 144~Si atoms. These dimensions allow a reasonable distance 
between dislocations, about $6|\mathbf{b}|\simeq23$~\AA\ ($\mathbf{b}$ is the Burgers vector, $|\mathbf{b}|=3.84$~\AA), while remaining 
affordable with ab initio computations \cite{Piz03PMA}. All atoms were relaxed, the imposed deformation being kept by the periodic conditions. 
In case of fixed boundaries, a $8\times16\times1$ cell, i.e. 128~Si atoms, was considered and dangling bonds belonging to edge atoms were saturated 
by hydrogens. The system was placed in the centre of a supercell whose dimensions allowed a 3~\AA\ vacuum left between a hydrogen and the  
supercell edge, and the imposed deformation was kept by relaxing only Si atoms not located on the system edges. The distance between the  
central dislocation and the frozen edges is then approximately 12~\AA.

\subsection{3. Results and discussion}

\begin{figure}[ht]
\includegraphics[width=8.2cm]{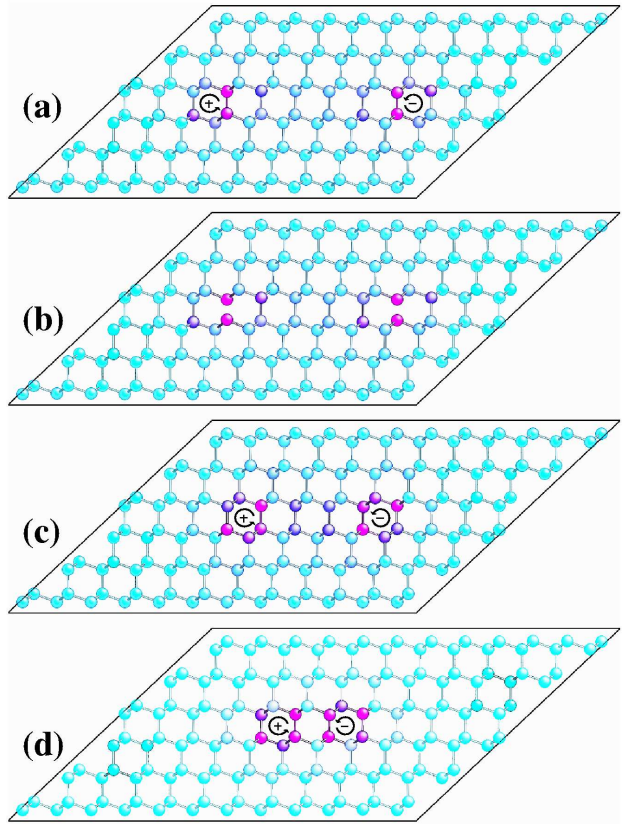}
\caption{Evolution of the system for a shear of 7\%: (a) initial configuration
of one dipole, with a distance $d=6|\mathbf{b}|$
(b) the two dislocations are approximately in the configuration B, with
$d\simeq5|\mathbf{b}|$ (c) each dislocation has moved by one hexagon, $d=
4|\mathbf{b}|$
(d) $d=2|\mathbf{b}|$. The colour of atoms indicates the magnitude of calculated
Von Mises stresses.}\label{Dismov}
\end{figure}

\noindent Starting with the periodic system, for shears up to 6\%, the dislocations remained in the vicinity of their initial locations. The energy of the relaxed system
increased with the deformation, due to the storage of elastic energy in the structure. However, when the applied deformation was equal to 7\%, the two dislocations of
the dipole started to move closer during the relaxation. The Figure~\ref{Dismov} shows different steps of the evolution of the structure. First, the bond on the long
edge of the hexagon broke, with dislocations approximately in the configuration B (b). Then the dislocations were found again in the configuration A, but in the next
hexagon (c). Finally, the dislocations continued to slip closer during the relaxation (d). In fact, there is an attractive force between the two dislocations with
opposite Burgers vectors, which increases as the dislocations come closer. This behaviour is different from what 
was obtained by Miyata et al., where the movement stopped once 
the dislocations slip by one hexagon \cite{Miy01PRB}. This difference may be explained by the different periodic boundary conditions used in their work, or by their
larger criterion value for stopping the relaxation ($10^{-3}$~Ry/bohr to compare with 
$10^{-4} $~Ry/bohr in our work). In the case of a fixed boundary system, we observed a
similar behaviour, the dislocation moving away from the centre, but only when the deformation is equal or greater than 8\%.

The stress required for moving the dislocations closer is simply the product of
the imposed shear with the relevant shear modulus $G$ \cite{noteG}.
So, in the case of periodic conditions, the calculated stress ranges between
$0.06G$ and $0.07G$.
However, due to periodic conditions, we have an infinite distribution of
dislocations, in interaction. This stress is then not
the real Peierls stress, due to these interactions. Considering that the energy
vary smoothly during the ABA move, and B is a saddle point, the dislocations are
expected to move once they have covered roughly half the distance between A and
B. Using isotropic elasticity theory, we have evaluated the extra stress at this
point to be about $-0.0051G$. The main contributions are coming from neighbor
dislocations along the slipping direction and favour the
slip, lowering the determined stress. The real Peierls stress then ranges between
$0.0651G$ and $0.0751G$, as determined with periodic conditions.
For fixed conditions, the stress ranges between $0.07G$ and $0.08G$. In that case too, the calculated stress is not the real Peierls 
stress. In fact, the fixed edges of the system create an extra force, opposed to the slip of the dislocation and increasing the calculated 
stress. Following the work of Shenoy and Phillips \cite{She97PMA}, we estimated the extra stress to be $0.0046G$ when the dislocation is 
located at midpoint between A and B. The Peierls stress then ranges between $0.0654G$ and $0.0754G$. The agreement between values calculated 
using periodic or fixed conditions is very good. Therefore we expect that our results are independent on the boundary conditions.
In the present situation where the Peierls stress is relatively large, the agreement between the two methods indicates that the use of sophisticated 
boundary treatments, such as flexible boundaries \cite{Rao98PMA}, is not essential. 

Our Peierls stress ranges between $0.0651G$ and $0.0754G$, i.e. between about $2.4\times10^{-2}$~eV~\AA$^{-3}$ and 
$2.8\times10^{-2}$~eV~\AA$^{-3}$. We have performed
atomistic simulations of the dislocation slip due to an applied stress, which is expected to be more precise than indirect methods, such as the Peierls-Nabarro model.
Nevertheless, if we compare with similar studies, we note several differences. Hence, Koizumi et al. determined a Peierls stress of 
$1.3\times10^{-2} $~eV~\AA$^{-3}$
\cite{Koi00PMA}, using the Stillinger-Weber potential. Our own tests, done in the same conditions, lead to a similar result, but also 
showed that this low value is due to
an incorrect description of the ABA path energy. In fact, the configuration B, weakly metastable with a metallic behaviour along the 
dislocation line \cite{Piz03PMA}, is the most stable with this potential. Ren et al. found a much larger value, 
$3.6\times10^{-2}$~eV~\AA$^{-3}$ \cite{Ren95PRB}, using Stillinger-Weber, but we
  have no explanation for this disagreement with 
Koizumi et al. and with our own tests. Finally, Miyata et al. have determined indirectly the Peierls stress from ab initio calculations and 
found a value between 14 and $19\times10^{-2}$~eV~\AA$^{-3}$ \cite{Miy01PRB}, much larger than our own value. Two possible explanations for 
such a disagreement can be found. The first one is related to the $10\times6\times2$ simulation cell they used. Along [121], dislocations in 
the cell were initially located at a $6|\mathbf{b}|$ distance, presumably giving $4|\mathbf{b}|$ between a dislocation and its periodic image. 
The dislocation distribution is then not truly quadrupolar. Also, along $[\bar{1}1\bar{1}]$, the distance between dislocations is only $3|\mathbf{b}|$. 
This leads to a strong interaction between dislocations, opposed to the dislocation slip. The second reason is related to the use of an elastic 
analytical model for estimating the Peierls stress from atomistic calculations. This unnecessary step is a possible source of errors.

In conclusion, we have performed an accurate determination of the Peierls stress
of a shuffle screw dislocation in silicon, using first principles and two
different kinds of boundary conditions.
The calculated value is between $2.4\times10^{-2}$ and $2.8\times10^{-2}
$~eV~\AA$^{-3}$, this range corresponding to the deformation increment of 1\%.
Since the uncertainty associated with the computational method is lower, we can
conclude that the Peierls stress is $(2.6\pm0.2)\times10^{-2}$~eV~\AA$^{-3}$.
This is slightly lower than the lowest experimental value \cite{expeSP}.
Regarding the plasticity of silicon, it has been shown that at low temperature
perfect shuffle dislocations are governing. The shear applied on the system
favoured the ABA path (Fig.~\ref{Shuffleglide}), so that the screw dislocation
remained in its shuffle plane. We also checked a possible cross-slip between shuffle and
glide planes by performing simulations with an applied
deformation of 10\% favouring the ACA path. A slip process
was obtained, the dislocations moving closer during the relaxation. However, the
ABA path was again selected, instead of ACA. Therefore, at 0 K, whatever the orientation of the applied shear stress, the undissociated 
screw keeps gliding in the shuffle set and does not spontaneously transform into a dissociated configuration in the glide set. Additional 
investigations have to be done to better understand the plasticity properties of silicon, and in particular the transition between shuffle 
mode at low temperature to glide mode at high temperature.

\subsection{Figures captions}

\begin{thebibliography}{28}
\bibitem[{not({\natexlab{a}})}]{notemu}
\bibinfo{note}{$\mu$ is the isotropic shear modulus; the average value for Si
  is 0.425~eV~\AA$^{-3}$ \cite{Hir82WIL}.}
\bibitem[{con()}]{configscrew}
\bibinfo{note}{Recent ab initio calculations have shown that the configuration
  A, where the screw dislocation is located in the center of one hexagon is
  most stable for Si \cite{Piz03PMA}. The configuration B, with the screw
  located on the long edge of one hexagon is weakly metastable. Another
  metastable configuration, C, is obtained for a screw located on the small
  edge of one hexagon.}
\bibitem[{not({\natexlab{b}})}]{noteG}
\bibinfo{note}{The $\langle110\rangle$\{111\} shear modulus $G$ is
  $1/3(C_{11}-C_{12}+C_{44})$ \cite{Hir82WIL} and is 0.366~eV \AA$^{-3}$ from
  our calculations (0.381~eV~\AA$^{-3}$ experimentally).}
\end{thebibliography}
\end{document}